\documentstyle[12pt,aasms4]{article}
\begin{document}

\title{Photometric Variability of Be/X-ray-Pulsar Binaries in the SMC } 

\author{ P.C.\ Schmidtke \& A.P.\ Cowley }

\affil{Dept.\ of Physics \& Astronomy, Arizona State Univ., Tempe, AZ,
85287-1504, USA; paul.schmidtke@asu.edu, anne.cowley@asu.edu } 

\begin{abstract}

We have studied the photometric variability of ten SMC Be/X-ray pulsars
using MACHO and OGLE-II data.  For some of these systems we have found
periodic behavior, including orbital outbursts and/or nonradial pulsations
(NRP) of the Be star.  For others we were unable to identify any clear
photometric periodicity, although their longterm light curves show
significant structure.  We present periodograms, phase dispersion
minimization (PDM) variances, and folded light curves for the systems
which exhibit periodic photometric variability. 

\end{abstract}

                                    
\section{Introduction}

The Be/neutron-star systems are an important subclass of high-mass X-ray
binaries.  These systems are particularly numerous in the Small Magellanic
Cloud where about four dozen are known (Coe et al.\ 2005).  Many are
transient X-ray sources, and a large fraction have been shown to be X-ray
pulsars (e.g. Coe et al.\ 2005, Liu et al.\ 2005, Haberl \& Pietsch 2004,
and references therein). 

The optical identifications for these sources depend on accurate X-ray
positions which have been greatly improved with data from {\it Chandra}
and {\it XMM-Newton}.  Two valuable sources of optical data for Magellanic
Cloud systems are the MACHO and OGLE-II surveys (e.g. Alcock et al.\ 1999,
Udalski et al.\ 1997, Zebrun et al.\ 2001, Szymanski 2005) that provide
regular photometric data in $V,R$ (MACHO) and $I$ (OGLE-II) over many
years.  For SMC sources the identification with Be stars is more secure if
they are also included in the catalogue of H$\alpha$ emission-line stars
of Meyssonnier \& Azzopardi (1993; hereafter designated as [MA93]). 

In previous papers we have studied the photometric properties of a sample
of Magellanic Cloud X-ray binaries (e.g. Schmidtke et al.\ 2004, Schmidtke
\& Cowley 2005a). In this paper we present results for ten more Be/X-ray
pulsars in the SMC.  The sources studied were selected both on the
basis of availability of longterm photometry and because they lack
previous optical studies.

\section{Analysis of Optical Data from the MACHO and OGLE-II Projects} 

As in our earlier work, we have relied on photometric data from the MACHO
and OGLE-II surveys.  We have transformed the `blue' and `red'
instrumental MACHO magnitudes into standard $V$ and $R$ colors (see Alcock
et al.\ 1999).  Each data set, or portion there of, was flattened
(prewhitened) to remove longterm changes in the mean brightness level by
fitting a low order polynomical to the data.  In sources which show large
longterm variations (ones we call ``swoopers") the light curve can only be
flattened in segments, and there are often parts that can't be flattened
at all because the slope of variation is too steep or irregular.  The
resulting light curve segments are then analyzed using the method of Horne
\& Baliunas (1986).  Often these systems show periodic ``outbursts'',
thought to be caused by interaction between the neutron star and the Be
star's disk near periastron passage.  In such cases the Horne \& Baliunas
method may not be the best way to search for periods, since it assumes
sinusoidal variations.  Thus, for some systems we also used the phase
dispersion minimization (PDM) technique described by Stellingwerf (1978).
This method can identify periods in a light curve whose shape is not
sinusoidal, but instead is characterized by recurring outbursts. 

Table 1 lists the X-ray sources studied and includes the MACHO coordinates
of the optical counterpart, the MACHO and OGLE-II identification numbers,
and other information about the systems.  In the text the sources we studied
are listed in order of right ascension for ease in locating the discussion 
about them.  Coe et al.\ (2005) have
introduced a useful identification number for Small Magellanic Cloud
pulsars which avoids the confusion of multiple X-ray names.  They use SXP
followed by the pulse period in seconds (e.g. SXP15.3 or SXP202), and they
give a convenient cross listing of names (their Table 1).  For all the
systems studied here we have included the SXP number.  In addition, Coe
presents finding charts for most of the SMC pulsars on his web
site\footnote{http://www.astro.soton.ac.uk/$\sim$mjc/smc}.  For the
systems we have studied, we have confirmed his identifications are the
same as ours. 

In an earlier paper on SMC pulsars (Schmidtke \& Cowley 2005a), we showed
that some Be/pulsars exhibit nonradial pulsations (NRP) with periods
typically less than a day.  Therefore, in all cases we have not only
searched for orbital periods, but also for pulsation periods.  Our
searches covered the period range from 0.2 to 1000 days using a variety of
techniques and subsamples of the data. 

Table 2 lists the periods found for some of the Be/pulsar systems.  For a
few systems possible periods are only given in the text, but not included
in the table, because we think they need further verification.  There are
other sources for which we found no clear photometric periods in spite of
detailed searches of the entire data set or subsections of it.  In the
section below we give details about the analysis of each source. 

\section{Individual Be/X-ray Pulsar Systems}

\subsection{ XMMU~J005152.2$-$731033 = RX~J0051.9$-$7311 = SXP172 }

This X-ray source was identified with a Be star in the SMC by Cowley et
al.\ (1997; their ``Star 1"; also [MA93]504).  Yokogawa et al.\ (2000)
later found this source (then named AX~J0051.6$-$7311) to show 172.4 s
X-ray pulsations.  Haberl \& Pietsch (2004) also detected the pulsar and
gave its pulse period as 172.21$\pm$0.13 s. 

The longterm OGLE-II light curve (Fig.\ 1) shows the source undergoes slow
dips and rises.  The MACHO data have several problems.  The west-pier $V$
data have many `dropouts' and hence are not usable.  $R$ data are only
available when the telescope was east of the pier.  Hence, the longterm
MACHO light curves are not continuous, making periods difficult to
identify.  When only east-pier data were studied, no clear periods were
found in either $V$ or $R$, and additionally none were found in any
subset of the MACHO data. 

However, OGLE-II $I$ data reveal a period of 69.9$\pm$0.6 days (Schmidtke
\& Cowley 2005b).  We assume this is the orbital period, although one
might have expected a somewhat longer orbital period based on Corbet's
(1984) P$_{pulse}$/P$_{orb}$ relation.  Fig.\ 1 shows the PDM and
periodogram, which are similar to those seen in some other Be/X-ray
systems.  Although the strongest peaks in the periodogram are at 1/2 and
1/3 the true period, the PDM variance is minimized at 69.9 days.  The
69.9-day folded light curve (Fig.\ 1) shows ``outbursts" lasting a few
tenths of the period, with an average amplitude of $\sim$0.02 mag. 
However, at some epochs the outbursts are more prominent.  In Fig.\ 1 we
have superimposed observations from the strongest outburst on the mean
folded light curve to show how much the amplitude varies.  We have also
searched for short periods (i.e. less than 2 days), but we found nothing
significant. 

Laycock et al.\ (2005) report a period of P$_X$=67$\pm$5 days based on
several X-ray outbursts.  This 67-day period is in acceptable agreement
with our optical period, considering their analysis is based on weekly
monitoring.  We note that their Table 6 lists a value of P(orb)=147$\pm$24
days, rather than the 67-day value given in their text. 

\subsection{ RX~J0051.8$-$7231 = SXP8.88 }

This variable X-ray source, also known as 1E0050.1$-$7247, was found to be
a pulsar by Israel et al.\ (1997).  Although there is some uncertainty
about its exact position, the optical counterpart is thought to be the Be
star [MA93]506 (which is also Star \#2 of Stevens et al.\ 1999).  Fig.\ 2
shows its longterm optical light curve from both MACHO and OGLE-II data.
Its irregular ``swooping" behavior continued during OGLE-III observations,
as shown by Coe et al.\ (2005).  They suggest that this source may be the
same as SXP16.6, since neither source has a well-defined position. 
However, Laycock et al.\ (2005) argue against this interpretation. 

From MACHO $R$ data, Coe et al.\ (2005) derived P$_{orb}$=185$\pm$4 days.
However, since the calibration of east- and west-pier data is not always
the same, this can introduce a half-year pseudo-period.  Hence, we have
processed data from the two pier configurations separately and merged the
data sets after prewhitening.  Using this technique, we did not find the
185-day period.  We suspect that it is an artifact of slightly different
calibrations for the east- and west-pier data. 

We have also subdivided both OGLE-II and MACHO photometry into time
segments.  In the segment near the start of OGLE-II observations (MJD
50600-50800) there is a weak periodicity at P=33.4$\pm$1.0 days.  This
appears to be caused by several small outbursts (perhaps near periastron
passage, with amplitudes varying from orbit to orbit).  These outbursts
are marked on the inset of Fig.\ 2.  Although these high points are from
several different orbital cycles, there is no evidence of this period in
any of the MACHO data.  New data would be needed to confirm this possible
periodicity.  Corbet et al.\ (2004) analyzed X-ray outbursts from this
source and found P$_X$=28.0$\pm$0.3 days, which is in general agreement
with the possible 33-day optical period. 

\subsection{ RX~J0052.1$-$7319 = SXP15.3} 

This transient source was found to be a pulsar by Finger et al.\ (2001).
It was optically identified with a Be star by Covino et al.\ (2001; their
``Star A" which is also [MA93]552).  The longterm MACHO and OGLE-II light
curves for this star are shown in Fig.\ 2.  In spite of a detailed
investigation of these data and various subsets of it, we were unable to
find any significant periods in $V$, $R$, or $I$. 

\subsection{ CXOU~J005323.8$-$722715 = SXP138  }

Using $Chandra$ data, Edge et al.\ (2004) found this source to be a 138-s
pulsar.  From the accurate X-ray position they identified the optical
counterpart as the Be star [MA93]667.  Edge et al.\ derived a photometric
(orbital) period of P$_{orb}$=125$\pm$1.5 days from MACHO data, although
details of their analysis are not given. 

The longterm $R$ light curve for this star is plotted in Fig.\ 2.  There
is a slow downward trend with occasional small ``outbursts'' (e.g. near MJD
50400 and 50620), but these are not periodic.  We have re-examined both
the $V$ and $R$ data, studying the east-pier and west-pier data
separately.  We find that there is a significant seasonal effect in which
the mean brightness from each pier differs.  Hence, it was necessary to
prewhiten the data from each pier configuration before combining them. 

We have searched for periods in both MACHO colors and in a variety of data
subsets.  Although the first half of the west-$R$ data shows a weak
periodicity near 122-123 days (perhaps the same period as reported by Edge
et al.), other time segments show different possible periods in the range
of 220-260 days.  It appears that each of these periods is related either
to the length of the seasonal data train in a pier configuration or gaps
between these data sets.  No single long period is consistent with all of
the photometry.  Each of these ``periods" produced a low-amplitude
sinusoidal light curve, rather than the outburst behavior usually
displayed in orbital periods. 

The periodograms for this source also show some power near $\sim$1 day,
but in each case these are aliases of the long ``periods'' discussed above.
We conclude that none of the peaks in the periodograms represents a true
orbital period, but rather they are artifacts of the data. 

\subsection{ XMMU~J005605.2$-$722200 = 2E0054.4$-$7237 = SXP140}

This 140-s pulsar was discovered by Sasaki, Pietsch, \& Haberl (2003)
using $XMM-Newton$ data.  They note that it may be the same source as
2E~0054.4$-$7237.  Its position is coincident with the Be star [MA93]904
whose longterm MACHO light curve is shown in Fig.\ 3.  The source is
clearly a ``swooper''.  Unfortunately it lies outside of the OGLE-II
fields, so $I$ photometry is not available.  Coe et al.\ (2005) note a
blending problem in the crowded field of this source (see image at:
~http://www.astro.soton.ac.uk/$\sim$mjc/smc/Finders/sxp140.jpg). 

The MACHO data were analyzed mainly in Seg A (MJD 49650-50500), since this
is the only part of the longterm light curve that could be prewhitened
reasonably well.  A period near 197$\pm$5 days is present in both $R$ and
$V$ photometry.  The folded light curve shows a typical orbital outburst,
with a rapid rise ($\sim$0.04 mag) and a more gradual decline.  In other
time segments of the longterm light curve, where it was impossible to
flatten the data, we examined the region where outbursts were expected. 
In most cases there was evidence for a brightening at the predicted times.
 The 197-day period is in reasonable agreement with Corbet's (1984)
P$_{pul}$/P$_{orb}$ relationship.  We conclude that this is probably the
orbital period since in similar systems the amplitude of outbursts varies
or may be absent. 

\subsection{ XMMU~J005920.8$-$722316 = SXP202 } 

This X-ray source was discovered to be a 202-s pulsar by Majid et al.\
(2004) using archival $XMM-Newton$ data.  Its X-ray position is coincident
with both an OGLE-II and a MACHO star.  The longterm light curve for this
system (Fig.\ 4) shows it to be another ``swooper''. 

We have searched the $I$, $R$, and $V$ data for possible periods.  The
photometric behavior is similar to SXP138, with various subsets of the
data showing weak power at periods related to the length of the data train
and its aliases.  In $I$ there is some power at $\sim$203 days and its
alias at $\sim$0.99 days, but light curves folded at these periods merely
bunch the observations into a few clumps and do not appear to be
reasonable.  We conclude that there are no clear periodities in the
photometric data.  Coe et al.\ (2005) came to the same conclusion. 

\subsection{ CXOU~J010102.7$-$720658 = RX~J0101.0$-$7206 = SXP304 }

Macomb et al.\ (2003) discovered this 304.5-s pulsar in $Chandra$
observations of the SMC.  Its position clearly identifies it with the Be
star [MA93]1240, which is the same star Edge \& Coe (2003) found to have
strong H$\alpha$ emission. 

Although no OGLE-II data are available, this star is included in the MACHO
data set.  The longterm MACHO $V$ light curve is shown in Fig.\ 5.  The
mean magnitude is relatively flat over a long stretch of time (MJD
49000-51000) allowing it to be prewhitened quite easily.  The periodogram
shows very prominent power at P=0.26 days, which we associate with
nonradial pulsations of the Be star.  When the data are subdivided into
four time segments, there is evidence that the period changes slightly
between segments A-B and C-D (as shown in Fig.\ 5).  The folded Seg C data
show that the amplitude is greatest in $V$ and least in $I$. 

We have also searched for longer periods which might arise from orbital
interactions between the pulsar and the Be star's disk.  Using the same
data in which the NRP were found, there is moderate power at P$\sim$520
days.  To further investigate this long period, we subtracted a sinusoid
with the period and amplitude of the NRP.  Analysis of the residuals still
reveals a periodicity at 520$\pm$12 days.  The light curve folded on this
period shows a steep rise and slower decline, typical of orbital
outbursts.  However, the significance of this period is low, and further
data are needed to confirm this possible orbital period. 

\subsection{ SAX~J0103.2$-$7209 = 2E~0101.5$-$7225 = SXP348 }

Israel et al.\ (2000) showed this X-ray source to be a 348-s pulsar.  Its
position identifies SXP348 with the Be star [MA93]904.  The MACHO data
have a significant number of `dropouts' and other problems, probably
related to its proximity to a much brighter star.  Although unlisted in
OGLE-II's Difference Image Analysis, photometry for the star is available
in the complete OGLE-II data set.  The $I$ light curve is shown in Fig.\
4. 

Analysis of the $I$ photometry shows a weak period at 93.9 days.  The
folded light curve shows a steep rise and slower decline, but additional
data would be needed to confirm this as the orbital period.  A system with
a 348-s pulse period would be expected to have an orbital period
considerably longer than 94 days. 

\subsection{ RX~J0103.6$-$7201 = SXP1323 }
 
Using $XMM-Newton$ data Haberl \& Pietsch (2005) discovered 1323-s X-ray
pulsations from this source, making it the slowest known SMC pulsar.  Its
optical counterpart is the Be star [MA93]1393.  Only OGLE-II data are
available for this star, but this photometry reveals a wealth of
information about the source's variability. 

The periodogram shows extremely significant power at P=26 days as well as
at other periods (see Fig.\ 6 and Schmidtke \& Cowley 2006).  When the
data are folded on the 26-day period, the variation is approximately
sinusoidal with a full amplitude of $\sim$0.04 mag.  From the known
relation between pulse period and orbital period (Corbet 1984), one would
expect the orbital period to be much longer than 26 days.  In addition,
orbital light curves usually show outbursts rather than sinusoidal
variations.  Hence, we conclude that the alias period at P=0.96 days (the
second strongest peak in the periodogram) is likely to be the true primary
period, with the variation due to nonradial pulsations of the Be star. 

The periodogram also shows many other peaks, some resulting from other
pulsational periods and others being aliases of the true periods.  The
next strongest peak is at P=0.88 days, which we identify as a second
pulsational mode of the Be star.  A third pulsational period is present at
0.42 days, as can be seen in Fig.\ 6. 

We further analyzed these periods using a variety of techniques.  Although
all three of the NRP periods given above can be seen in the original
periodogram, we experimented with the data by subtracting the 0.96-d
period and examining the periodogram of the residuals.  This showed the
0.88-day period to be highly significant.  The 0.96-day light curve must
be a nearly perfect sine curve since no evidence of it was found in the
residuals.  We then subtracted the 0.88-day period, and examined the new
residuals.  This showed periods at both 0.42 and 0.28 days (aliases of
each other), with the 0.42-day period being slightly more significant.
Finally, we note that when all three NRPs (0.96, 0.88, and 0.42 days) are
subtracted from the data, the resulting residuals show another weak peak
at 1.05 days, suggesting that a fourth NRP period may be present.  Thus,
this source has many pulsational modes, all present at the same time.  We
also looked for changes in these periods by subdividing the longterm light
curve into 4 time segments.  No significant changes in any of the periods
were found. 

\subsection{ RX~J0105.1$-$7211 = AX~J0105$-$722 = SXP3.34 }

This 3.34-s pulsar was identified by Coe et al.\ (2005) with the Be star
[MA93]1506.  The longterm MACHO light (Fig.\ 7) shows a fairly steady mean
light level with considerable scatter.  Coe et al.\ found an optical
period of 11.09 days from these data, but he pointed out that the strength
of H$\alpha$ emission (EW=$-$54\AA) was not consistent with the
P$_{orb}$/EW relationship of Reig, Fabrigat, \& Coe (1997).  We have
searched to higher frequencies and find that the fundamental period is at
1.099 days, with the 11-day period being an alias of that period
(Schmidtke \& Cowley 2005c).  The power at the 1.009-day period is
exceptionally strong (Fig.\ 7).  This period is extremely stable in both
frequency and amplitude.  The folded $V$, $B$, and $I$ light curves show
the greatest amplitude in $I$ and the smallest in $V$, although the
differences are small.  This short period clearly is another example of
prominent nonradial pulsations from the Be star. 

When the 1.099-day variation is subtracted from the data, analysis of the
residuals shows a periodogram with a second significant peak at P=0.980
days.  The folded light curve reveals a very low-amplitude sinusoid in
each color (Fig.\ 7).  When the data are divided into 4 time segments, we
find that the amplitude is weakest in the last segment (starting about MJD
50950). 

\section{ Summary}

We have looked for general conclusions regarding the Be/X-ray pulsars
systems based on their photometric behavior.  We have considered not only
the sources in this paper, but also results from previous studies (e.g.
Coe et al.\ 2005, Schmidtke \& Cowley 2005a, Schmidtke et al.\ 2004).  We
see no correlation between the length of the X-ray pulse period and
whether or not orbital periods are seen in the photometric data.  Sources
with short pulse periods (e.g. 7.8 s for SMC X-3), long pulse periods
(e.g. SXP755 with P$_{orb}$=394 days), and many intermediate values may or
may not show orbital outbursts.  The presence of such outbursts must
depend on the orbital eccentricity and the extent of the Be star's disk at
the time of periastron passage.  Perhaps the systems which do not show
outbursts have nearly circular orbits. 

We also see that NRP are found in sources with a wide range of X-ray pulse
periods (from 3.34 s to 1323 s).  However, we note that sources with
``swooping'' light curves generally do not show NRP, even in the flattest
portions of their light curves (e.g. SXP8.88).  The only possible
exception to date is SXP82.4 whose OGLE-III light curve (Coe et al.\ 2005)
fades considerably, suggesting it may be a ``moderate swooper".  Its
1.33-day NRP were found only in the early part of the longterm light curve
when its the mean brightness was almost level (see Fig.\ 4 in Schmidtke \&
Cowley 2005a). 

In summary, we have analyzed the longterm optical light curves of ten SMC
Be/X-ray pulsars.  We determined orbital periods for four sources and a
possible orbital period for another system.  Overall, the orbital periods
found in this study fit reasonably well with the Corbet relation between
orbital and X-ray pulse period.  We have identified strong nonradial
pulsations for three sources, and two of these sources show multiple NRP. 

\acknowledgments

This paper utilizes public domain data obtained by the MACHO Project,
jointly funded by the US Department of Energy through the University of
California, Lawrence Livermore National Laboratory under contract No.\
W-7405-Eng-48, by the National Science Foundation through the Center for
Particle Astrophysics of the University of California under coopertative
agreement AST-8809616, and by the Mount Stromlo and Siding Spring
Observatory, part of the Australian National University. 

The OGLE-II database, as described by Udalski et al.\ (1997), Zebrun et
al.\ (2001), and Szymanski (2005), was also extensively used for this
project.

\newpage

\begin{figure}
\caption{ (top) Longterm $I$ light curve of XMMU~J005152.2$-$731033
(SXP172).  ~(middle) Variance calculated by the PDM method (upper, with
scale on right) and periodogram (lower) of $I^*$ data from Seg A. The 99\%
confidence level is also shown.  ~(bottom) $I^*$ data folded on the
69.9-day period and binned (filled squares), with the individual points
from the outburst near MJD 50790 superimposed (open circles). } 
\end{figure}

\begin{figure}
\caption{ (top) Longterm $V$ (upper) and $I$ (lower) light curves of
RX~J0051.8$-$7231 (SXP8.88).  The inset shows an enlargement of the region
between MJD 50600 and 50800, with four outbursts marked.  ~(middle)
Longterm $R$ (upper) and $I$ (lower) light curves of RX~J0052.1$-$7319
(SXP15.3).  ~(bottom) Longterm $R$ light curve of CXOU~J005323.8$-$722715
(SXP138).  } 
\end{figure}

\begin{figure}
\caption{ (top) Longterm $R$ light curve XMMU~J005605.2$-$722200
(SXP140).  ~(middle) Periodogram of $R^*$ data from Seg A showing the
orbital period of 197 days.  ~(bottom) $R^*$ light curve folded on 197
days and averaged in 20 phase bins.  } 
\end{figure}

\begin{figure}
\caption{(top) Longterm $R$ (upper) and $I$ (lower) light curves of 
XMMU~J005920.8$-$722316 (SXP202).  ~(bottom) Longterm $I$ light curve of
SAX~J0103.2$-$7209 (SXP348).} 
\end{figure}

\begin{figure}
\caption{ (top) Longterm $V$ (upper) and $I$ (lower) light curves of 
CXOU~J010102.7$-$720658 (SXP304).  ~(middle) Periodograms of $V^*$ data
from 4 time segments (A -- D) showing the variation in period and power.
~(bottom) Folded light curve of $V^*$, $R^*$, and $I^*$ data from Seg C
showing strong nonradial pulsations at P=0.26 days.  The amplitude appears
to be greatest in $V^*$.  } 
\end{figure}

\newpage

\begin{figure}
\caption{ (top) Longterm $I$ light curve of RX~J0103.6$-$7201
(SXP1323).  ~(second) Periodogram of $I^*$ data showing several
significant periods and their aliases (see text). ~(third) $I^*$ data,
folded on P=0.96 days, after signals from P=0.88 and P=0.41 days have been
subtracted.  ~(fourth) $I^*$ data, folded on P=0.88 days after the
variation of the other two NRP have been subtracted.  ~(bottom) Similar to
the above two plots, but the $I^*$ data have been folded on P=0.41 days
after subtraction of the variation due to the two other short periods. } 
\end{figure}

\begin{figure}
\caption{ (top) Longterm $R$ (upper) and $I$ (lower) light curves of 
RX~J0105.1$-$7211 (SXP3.34).  ~(second) Periodogram of $R^*$ data showing
the exceptionally strong peak at P=1.099 days.  ~(third) Folded and binned
light curves in $V^*$, $R^*$, and $I^*$ at P=1.099 days.  The amplitude is
largest in $I^*$.  ~(fourth) Periodogram of $R^*$ data after the 1.099-day
variation has been subtracted.  The strongest remaining power is at
P=0.980 days.  ~(bottom) $V^*$, $R^*$, and $I^*$ light curves, folded on
P=0.980 days, after removal of the 1.099-day signal.  See text for further
discussion of this source.} 
\end{figure}

\newpage

\begin{deluxetable}{llll}
\tablenum{1}
\tablecaption{Small Magellanic Cloud Be/Neutron-Star Systems Studied}
\tablehead{  
\colhead{System Name} &
\colhead{R.A.(2000)\tablenotemark{a}} &
\colhead{Dec.(2000)\tablenotemark{a}} &
\colhead{SXP \#} \\
\colhead{~~~~~~~OGLE-II \#} &
\colhead{~~~~~MACHO \#} &
\colhead{~~~~~~~[MA93] \#\tablenotemark{b}} & 
\colhead{ }
}
\startdata
XMMU~J005152.2$-$731033 &  00:51:51.86 & $-$73:10:33.4 &SXP172  \nl
  ~~~~~005152.02$-$731033.7 &~~~~~212.16077.13 & ~~~~~~~~504 & \nl
RX~J0051.8$-$7231 &  00:51:53.00 & $-$72:31:48.1 & SXP8.88 \nl
 ~~~~~005153.12$-$723148.3 & ~~~~~208.16087.9 & ~~~~~~~~506 & \nl
RX~J0052.1$-$7319 & 00:52:13.98 & $-$73:19:18.1  & SXP15.3\nl
 ~~~~~005213.99$-$731918.3 & ~~~~~212.16075.13 & ~~~~~~~~552  & \nl
CXOU~J005323.8$-$722715 & 00:53:23.86 & $-$72:27:15.1 & SXP138  \nl
 ~~~~~~~~~~- - - - -  &~~~~~207.16202.50 & ~~~~~~~~667 & \nl
XMMU~J005605.2$-$722200 & 00:56:05.62 & $-$72:22:00.2  & SXP140 \nl
 ~~~~~~~~~~- - - - -  &~~~~~207.16374.21 & ~~~~~~~~904 & \nl
XMMU~J005920.8$-$722316 & 00:59:20.94 & $-$72:23:17.4 & SXP202  \nl
 ~~~~~005921.03$-$722317.1 &~~~~~207.16545.12 & ~~~~~~~~--- & \nl
CXOU~J010102.7$-$720658 & 01:01:02.56 & $-$72:06:56.5  & SXP304 \nl
 ~~~~~010102.87$-$720658.7  & ~~~~~206.16663.16 &~~~~~~~~1240 & \nl
SAX~J0103.2$-$7209 & 01:03:13.93 & $-$72:09:13.8 & SXP348 \nl
 ~~~~~010313.89$-$720914.0 &~~~~~206.16776.17 & ~~~~~~~~904 & \nl
RX~J0103.6$-$7201 & 01:03:37.50 & $-$72:01:32.9  &  SXP1323 \nl
 ~~~~~010337.50$-$720132.9  & ~~~~~- - - - - &~~~~~~~~1393 & \nl
RX~J0105.1$-$7211 & 01:05:02.40 & $-$72:10:55.9  & SXP3.34  \nl
 ~~~~~010502.51$-$721055.9  & ~~~~~206.16890.17 &~~~~~~~~1506 & \nl
\enddata

\tablenotetext{a}{Coordinates are the MACHO position of the optical star,
except for RX~J0103.6$-$7201. }

\tablenotetext{b}{From catalogue of SMC H$\alpha$ emission-line stars
by Meyssonnier \& Azzopardi (1993). } 

\end{deluxetable}

\newpage

\begin{deluxetable}{lccl}
\tablenum{2}
\tablecaption{Photometric Periods for 
SMC Be/Neutron-Star Systems}
\tablehead{  
\colhead{System Names} &
\colhead{P$_{orb}$} &
\colhead{NRP\tablenotemark{a}} &
\colhead{References} \\
\colhead{} &
\colhead{(days)} & 
\colhead{(days)} & 
\colhead{}
}
\startdata
XMMU~J005152.2$-$731033 (SXP172)\tablenotemark{b} & 69.9$\pm$0.6 & --- & 
Laycock et al.\tablenotemark{c} \nl
RX~J0051.8$-$7231 (SXP8.88) & 33.4$\pm$1.0: & --- & Corbet et 
al.\tablenotemark{d} \nl
XMMU~J005605.2$-$722200 (SXP140) & 197$\pm$5: & --- & \nl 
CXOU~J010102.7$-$720658 (SXP304) &  520$\pm$12: & 0.26 & Schmidtke \& 
Cowley\tablenotemark{e} \nl
RX~J0103.6$-$7201 (SXP1323) & --- & 0.98; 0.88; 0.41 & \nl
RX~J0105.1$-$7211 (SXP3.34) & --- & 1.099; 0.980 & Coe et 
al.\tablenotemark{f} \nl
\enddata

\tablenotetext{a}{NRP = nonradial pulsations of the Be star. } 

\tablenotetext{b}{Optical counterpart is Star 1 in field of SMC~25 
(Cowley et al.\ 1997). }

\tablenotetext{c}{Laycock et al.\ (2005) find P$_{orb}$=67$\pm$5 days from
X-ray outbursts.}

\tablenotetext{d}{Corbet et al.\ (2004) give P$_{orb}$=28.0$\pm$0.3 days from
X-ray outbursts.}

\tablenotetext{e}{Schmidtke \& Cowley (2006) give further information about
this source.}

\tablenotetext{f}{Coe et al.\ (2005) give P$_{orb}$=11.09 days, which
is an alias of the strong NRP at 1.099 days (Schmidtke \& Cowley 2005c).} 

\end{deluxetable}

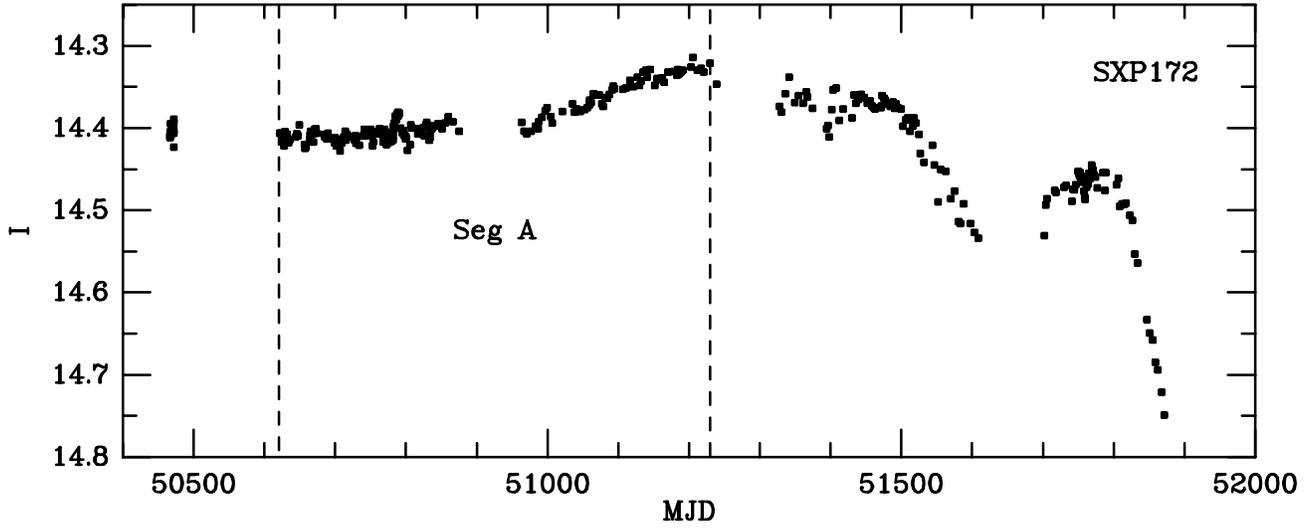
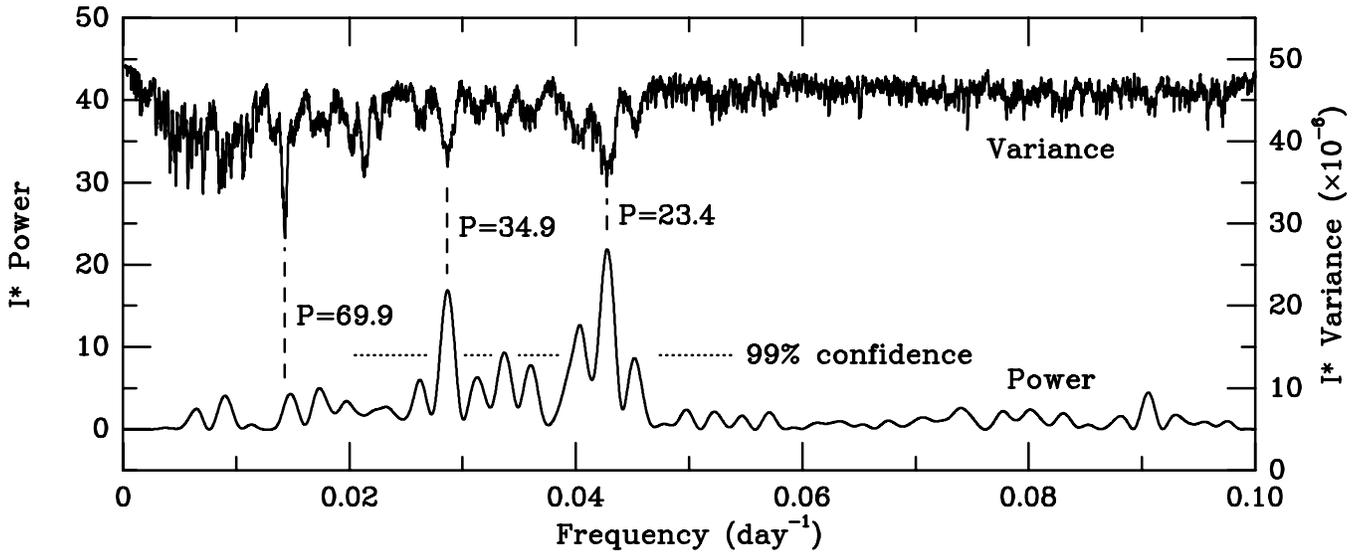
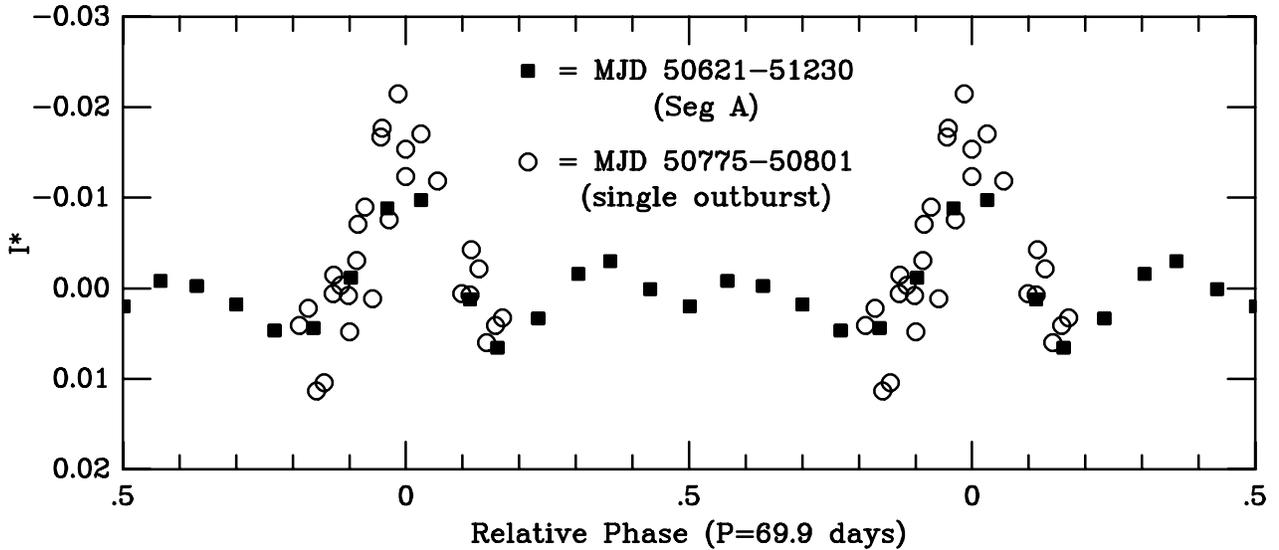
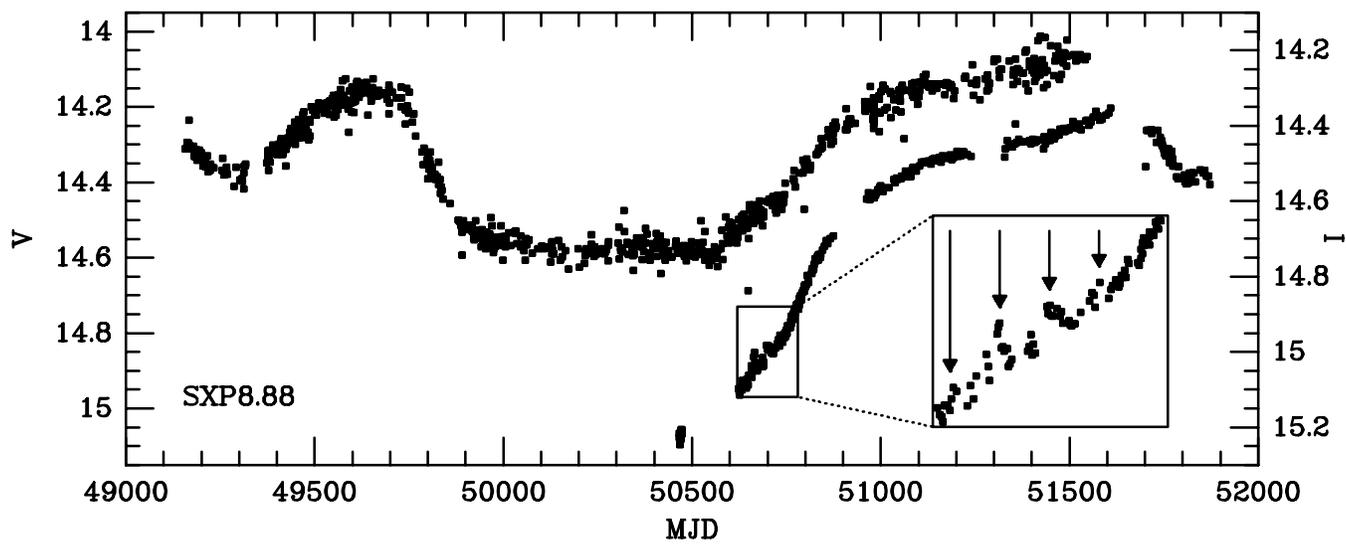
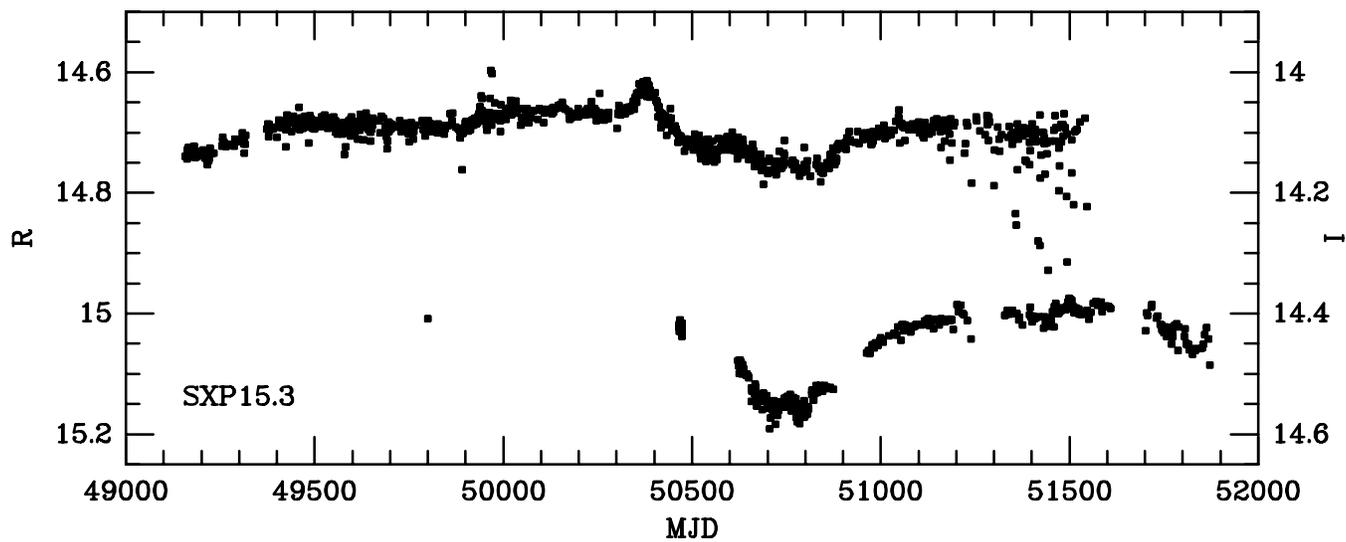
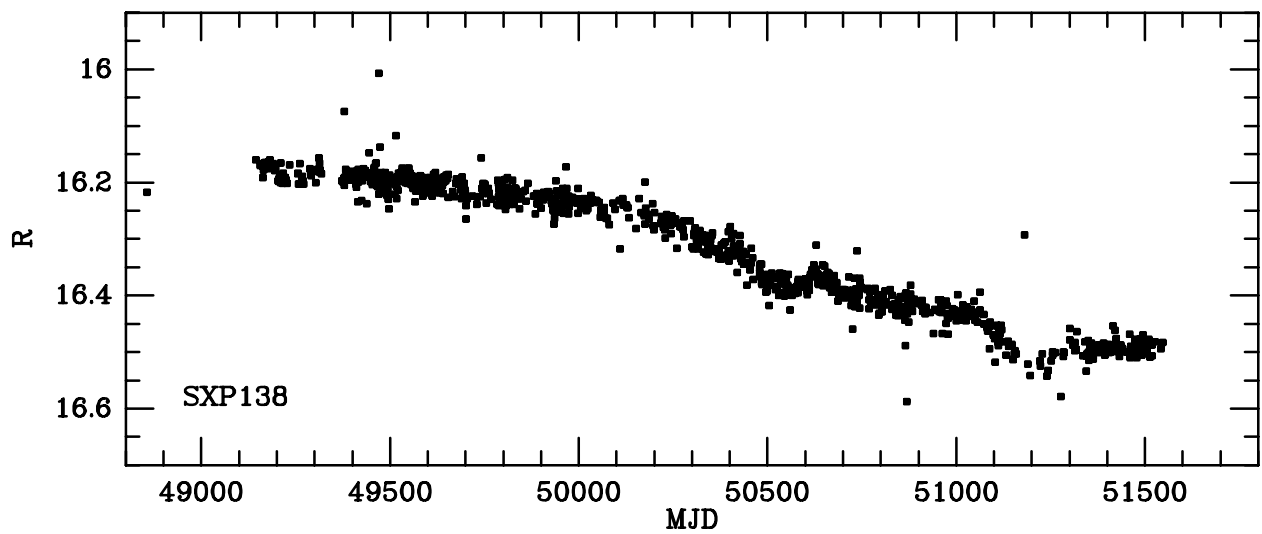
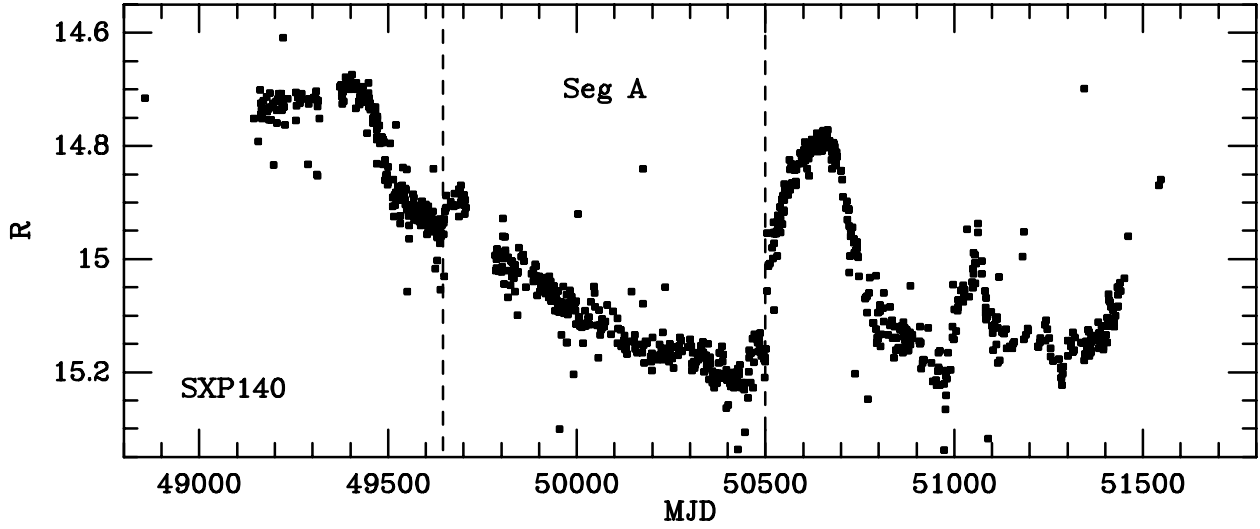
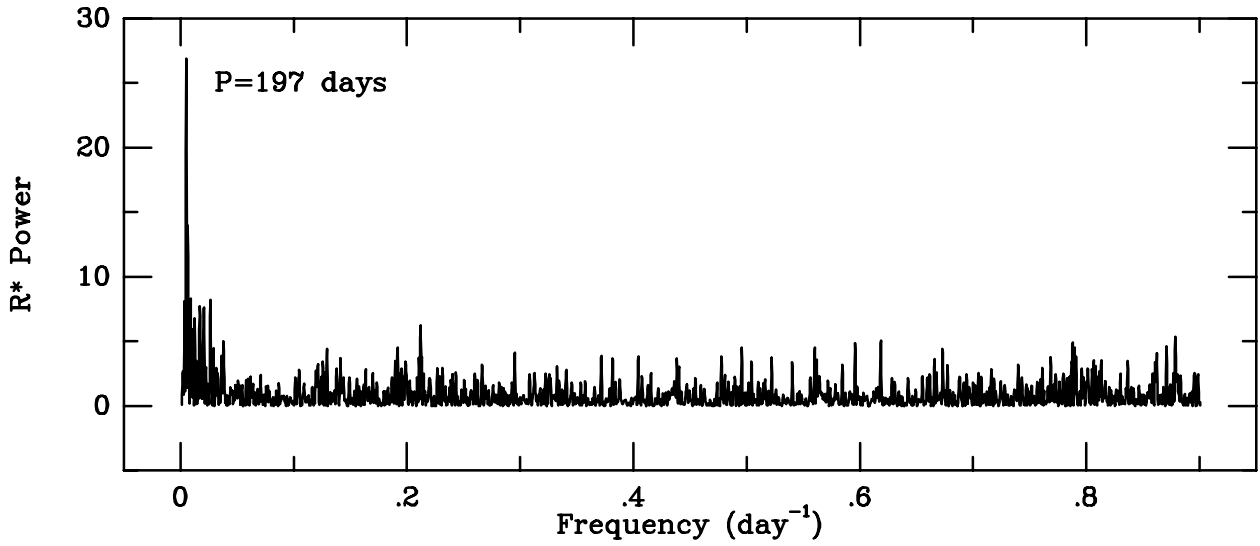
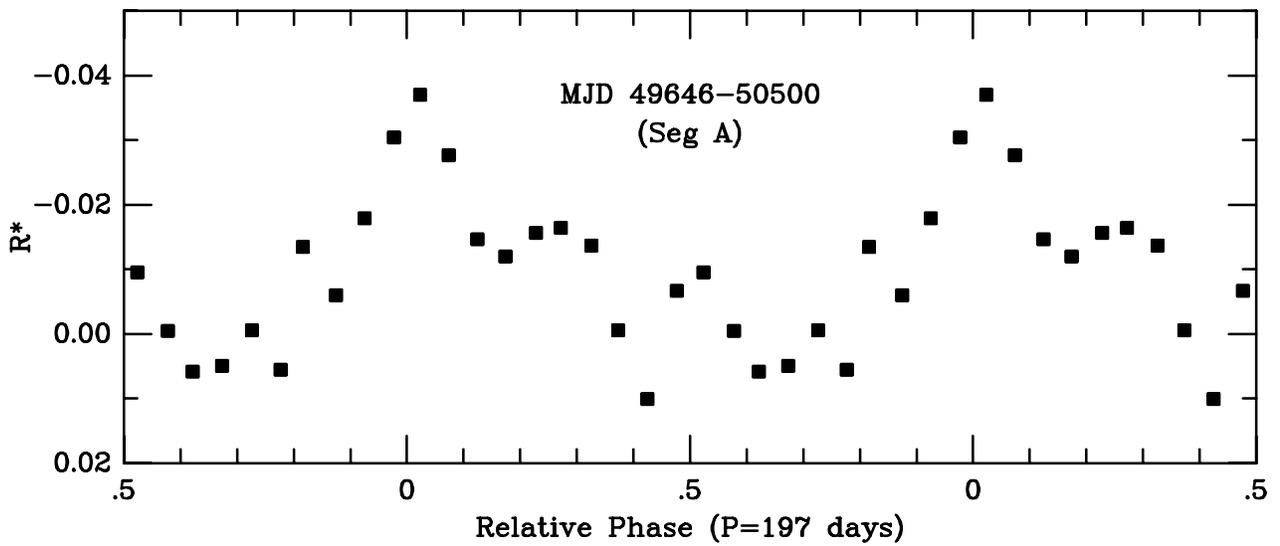
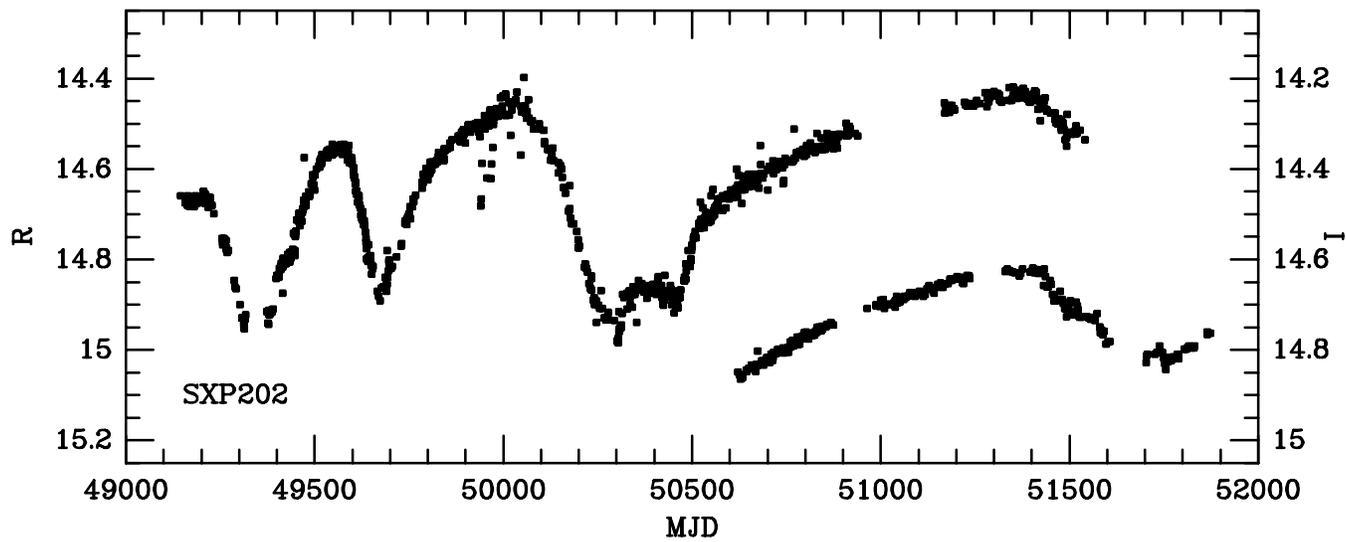
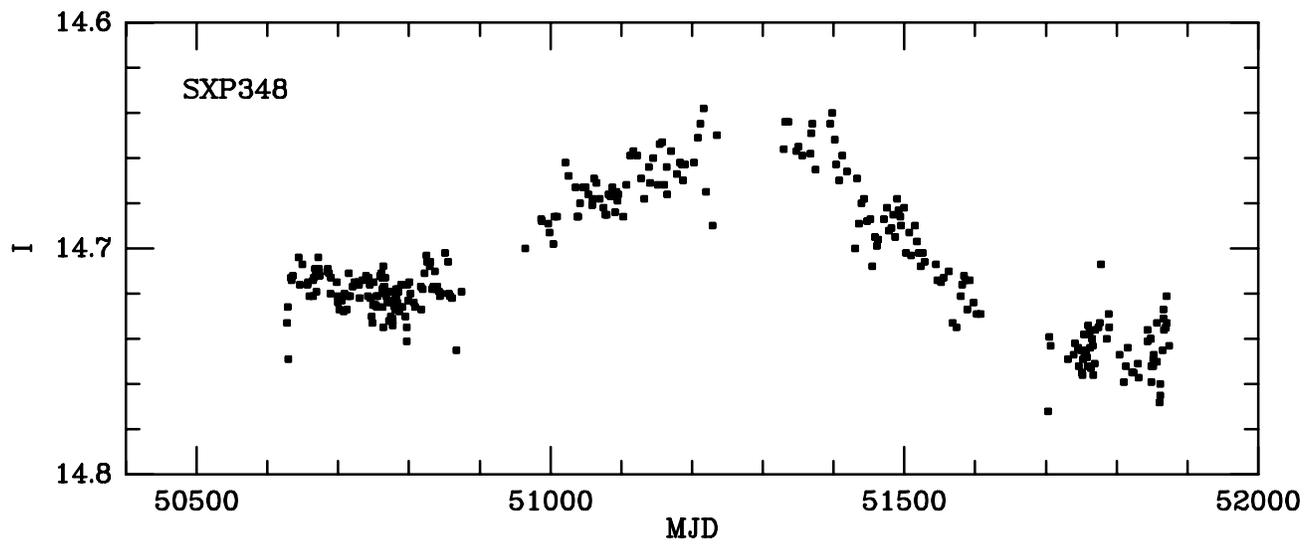
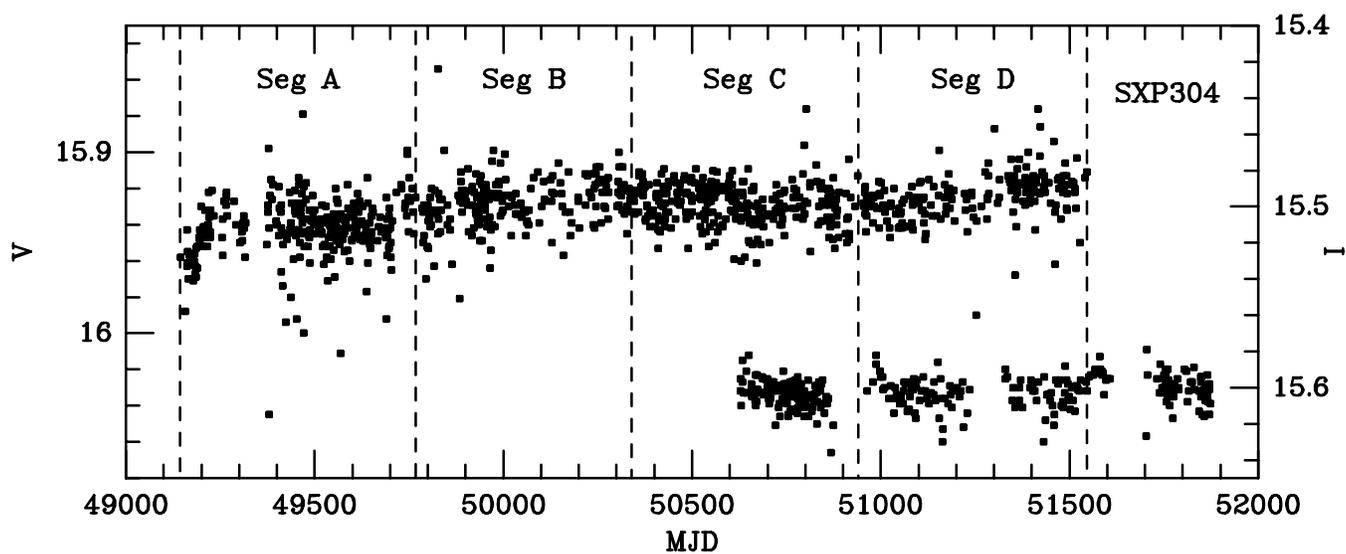
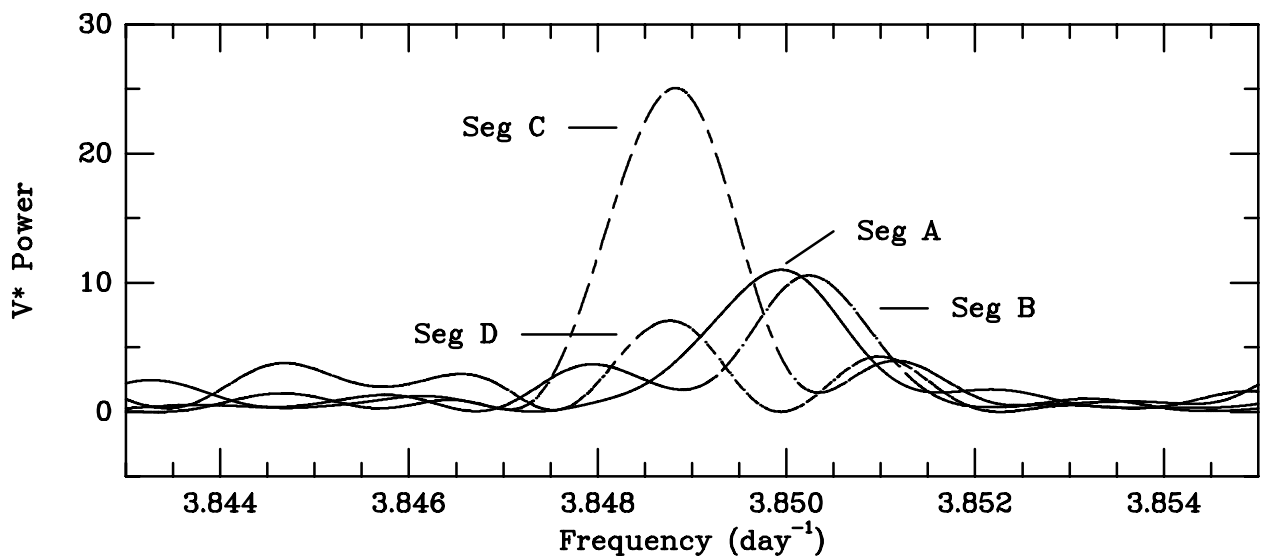
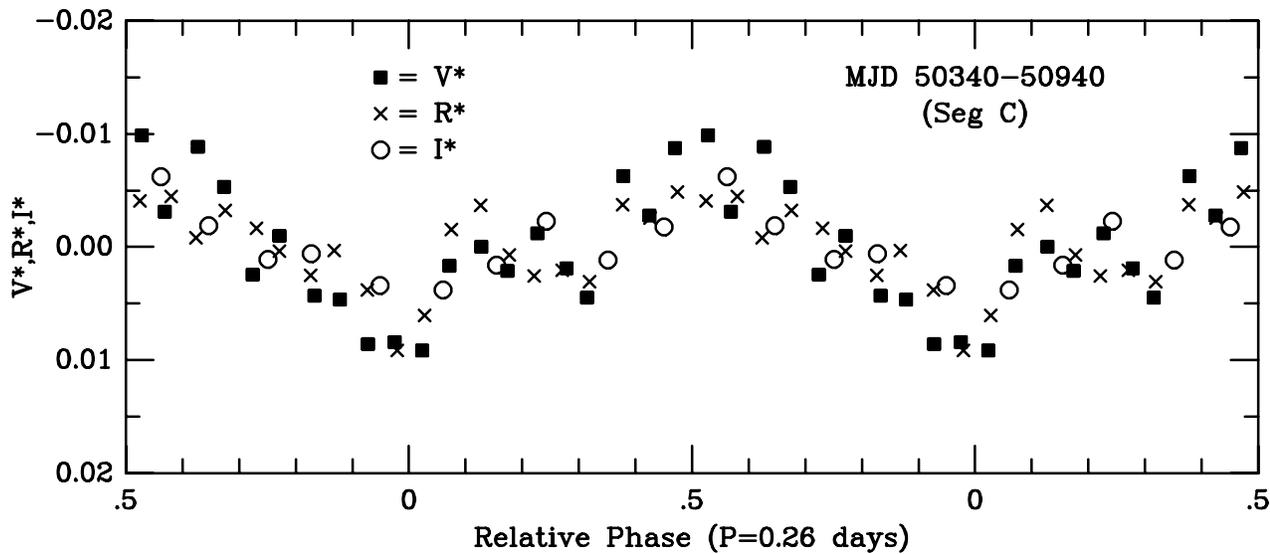
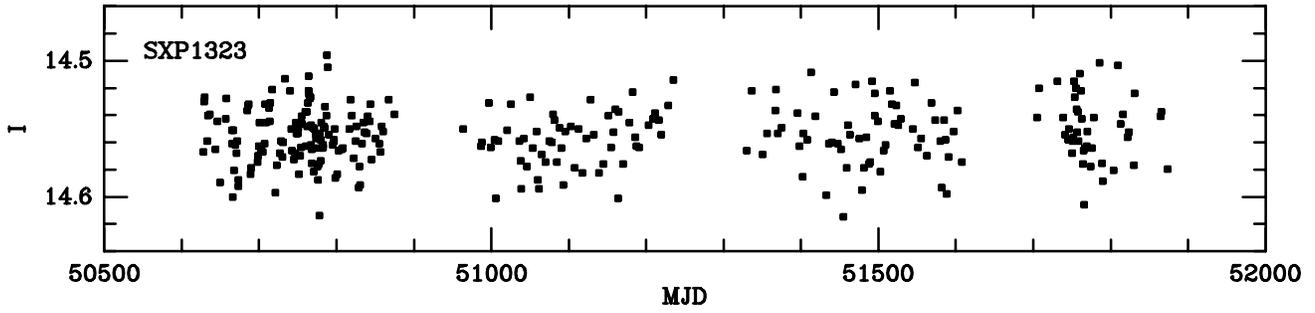
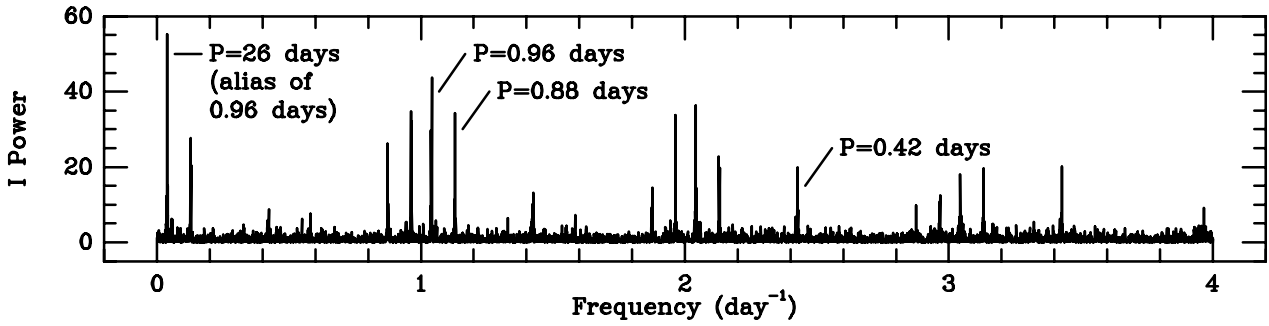
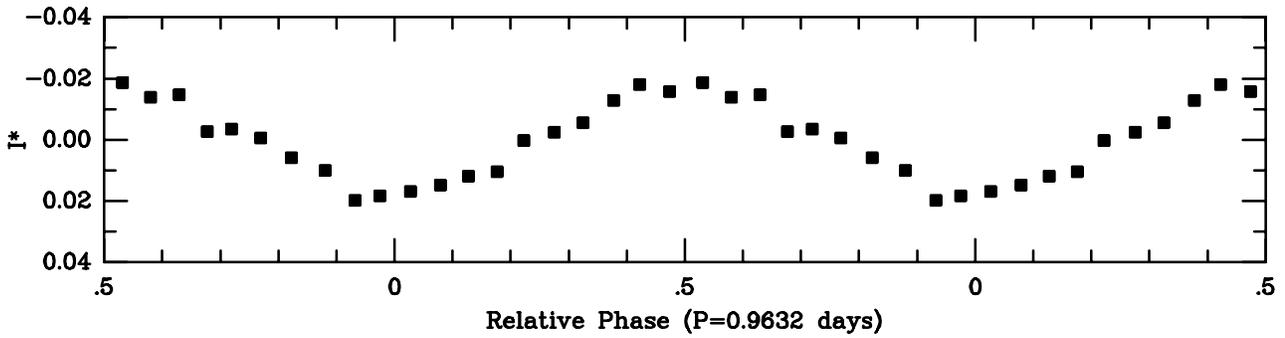
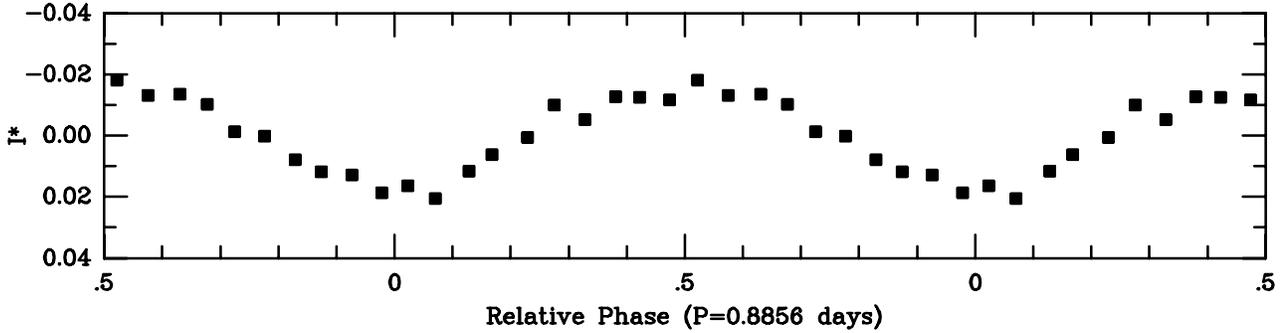
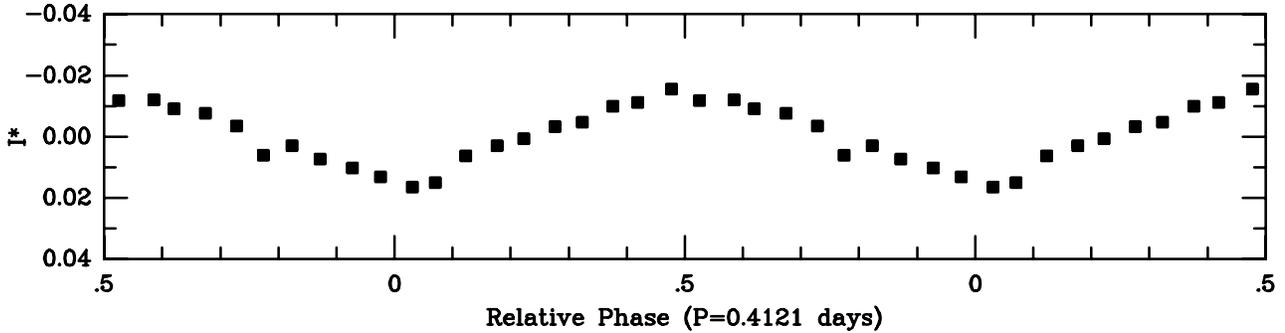
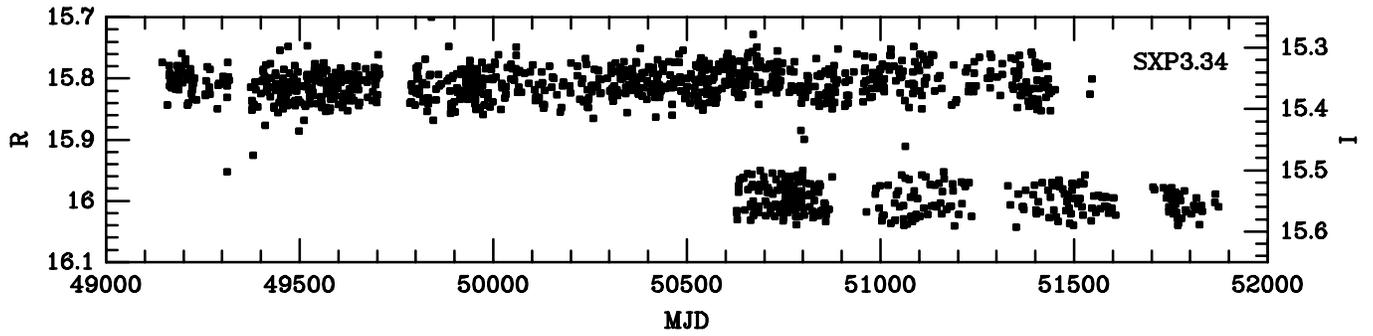
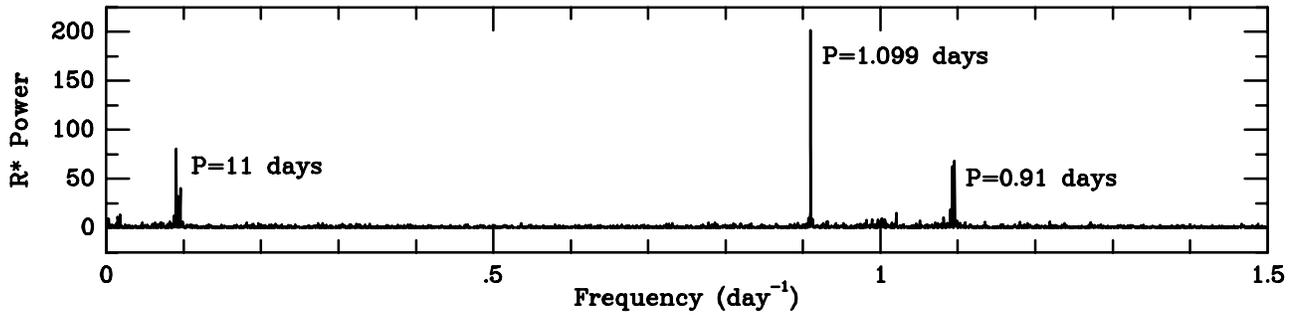
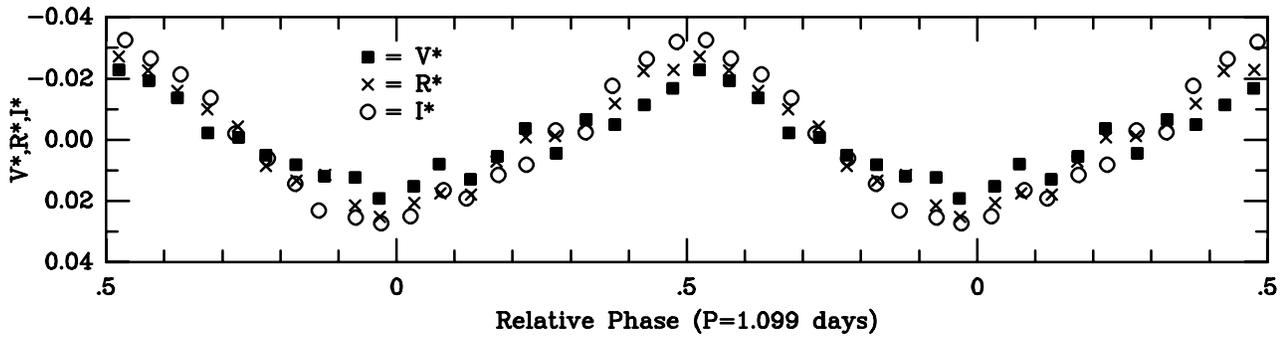
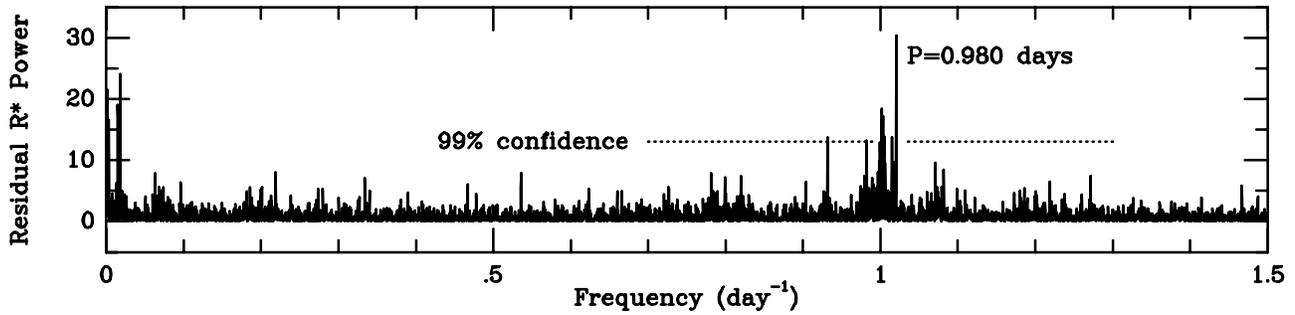
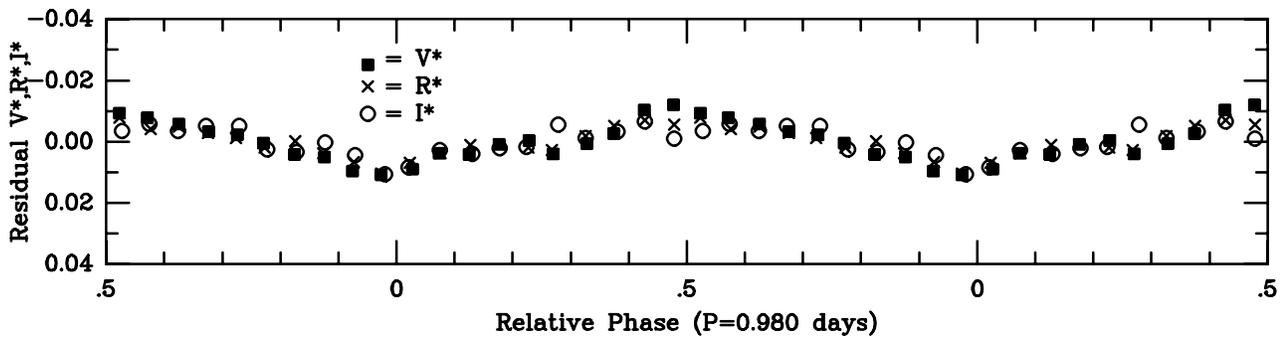
\end{document}